\documentclass[conference]{IEEEtran}
\IEEEoverridecommandlockouts
\usepackage{cite}
\usepackage{amsmath,amssymb,amsfonts}
\usepackage{algorithmic}
\usepackage{graphicx}
\usepackage{textcomp}
\usepackage{xcolor}
\def\BibTeX{{\rm B\kern-.05em{\sc i\kern-.025em b}\kern-.08em
    T\kern-.1667em\lower.7ex\hbox{E}\kern-.125emX}}

\usepackage{hyperref}
\usepackage{cleveref}
\usepackage{caption}
\usepackage{subcaption}
\usepackage{paralist}
\usepackage{tabularx,booktabs,longtable}
\usepackage{algorithm}
\usepackage{algorithmic}

\usepackage{listings}
\usepackage{color}
\usepackage{verbatim}

\definecolor{mygray}{rgb}{0.95,0.95,0.95}
\begin{document}

\title{Workflow-Driven Modeling for the Compute Continuum: An Optimization Approach to Automated System and Workload Scheduling\\
}

\author{\IEEEauthorblockN{1\textsuperscript{st} Aasish Kumar Sharma}
\IEEEauthorblockA{\textit{Faculty of Mathematics and Computer Science} \\
\textit{Georg-August-Universität Göttingen}\\
Göttingen, Germany \\
0000-0002-7514-2340}
\and
\IEEEauthorblockN{2\textsuperscript{nd} Christian Boehme}
\IEEEauthorblockA{\textit{Work Group Computing} \\
\textit{GWDG}\\
Göttingen, Germany \\
christian.boehme@gwdg.de}
\and
\IEEEauthorblockN{3\textsuperscript{rd} Patrick Gelß}
\IEEEauthorblockA{\textit{Department of Mathematics} \\
\textit{Zuse Institute, Berlin, Germany}\\
Berlin, Germany \\
gelss@zib.de}
\and
\IEEEauthorblockN{4\textsuperscript{th} Ramin Yahyapour}
\IEEEauthorblockA{\textit{Faculty of Mathematics and Computer Science} \\
\textit{Georg-August-Universität Göttingen}\\
Göttingen, Germany \\
ramin.yahyapour@gwdg.de}
\and
\IEEEauthorblockN{5\textsuperscript{th} Julian Kunkel}
\IEEEauthorblockA{\textit{Faculty of Mathematics and Computer Science} \\
\textit{Georg-August-Universität Göttingen}\\
Göttingen, Germany \\
julian.kunkel@gwdg.de}
}

\maketitle
\thispagestyle{plain}
\pagestyle{plain}

\begin{abstract}

The convergence of IoT, Edge, Cloud, and HPC technologies creates a compute continuum that merges cloud scalability and flexibility with HPC’s computational power and specialized optimizations. However, integrating cloud and HPC resources often introduces latency and communication overhead, which can hinder the performance of tightly coupled parallel applications. Additionally, achieving seamless interoperability between cloud and on-premises HPC systems requires advanced scheduling, resource management, and data transfer protocols. Consequently, users must manually allocate complex workloads across heterogeneous resources, leading to suboptimal task placement and reduced efficiency due to the absence of an automated scheduling mechanism.

To overcome these challenges, we introduce a comprehensive framework based on rigorous system and workload modeling for the compute continuum. Our method employs established tools and techniques to optimize workload mapping and scheduling, enabling the automatic orchestration of tasks across both cloud and HPC infrastructures. Experimental evaluations reveal that our approach could optimally improve scheduling efficiency, reducing execution times, and enhancing resource utilization. Specifically, our MILP-based solution achieves optimal scheduling and makespan for small-scale workflows, while heuristic methods offer up to 99\% faster estimations for large-scale workflows, albeit with a 5–10\% deviation from optimal results. Our primary contribution is a robust system and workload modeling framework that addresses critical gaps in existing tools, paving the way for fully automated orchestration in HPC-compute continuum environments.

\end{abstract}


\section{Introduction}
\label{sec:Introduction}

High-Performance Computing (HPC) systems are akin to the human body, where diverse components work in unison. Although the human body is naturally heterogeneous, HPC systems have traditionally been uniform; however, increasing variability now impacts performance, efficiency, and costs. Our goal is to illustrate a HPC-compute continuum that users can access without deep technical details. To this end, we first outline the system \& workload models, identify existing gaps, and then present tools \& techniques for mapping \& scheduling heterogeneous resources to optimize their utilization.

The article is organized as follows: Introduction (\Cref{sec:Introduction}) \& (\Cref{sec:Background}) provides a topic overview and the background on relevant system and workload context. Literature Review (\Cref{sec:RelatedWork}) surveys related work. Research Methodology (\Cref{sec:ResearchMethodology}) describes our approach. System and Workload Modeling (\Cref{ssec:SystemAndWorkloadModeling}) details the modeling, mapping, and scheduling strategies. Validation and Experiments (\Cref{sec:ObservationAndDiscussion}) present our findings. Conclusion and Recommendations (\Cref{sec:ConclusionAndRecommendations}) summarize the study and suggest future research directions. Additional sections include references and acknowledgments.

\vspace{-.45em}

\section{Background}
\label{sec:Background}

\subsection{System: HPC Compute Continuum}
\label{SystemHPCCompute Continuum}
\begin{figure}[ht]
    \centering
    \includegraphics[width=0.48\textwidth]{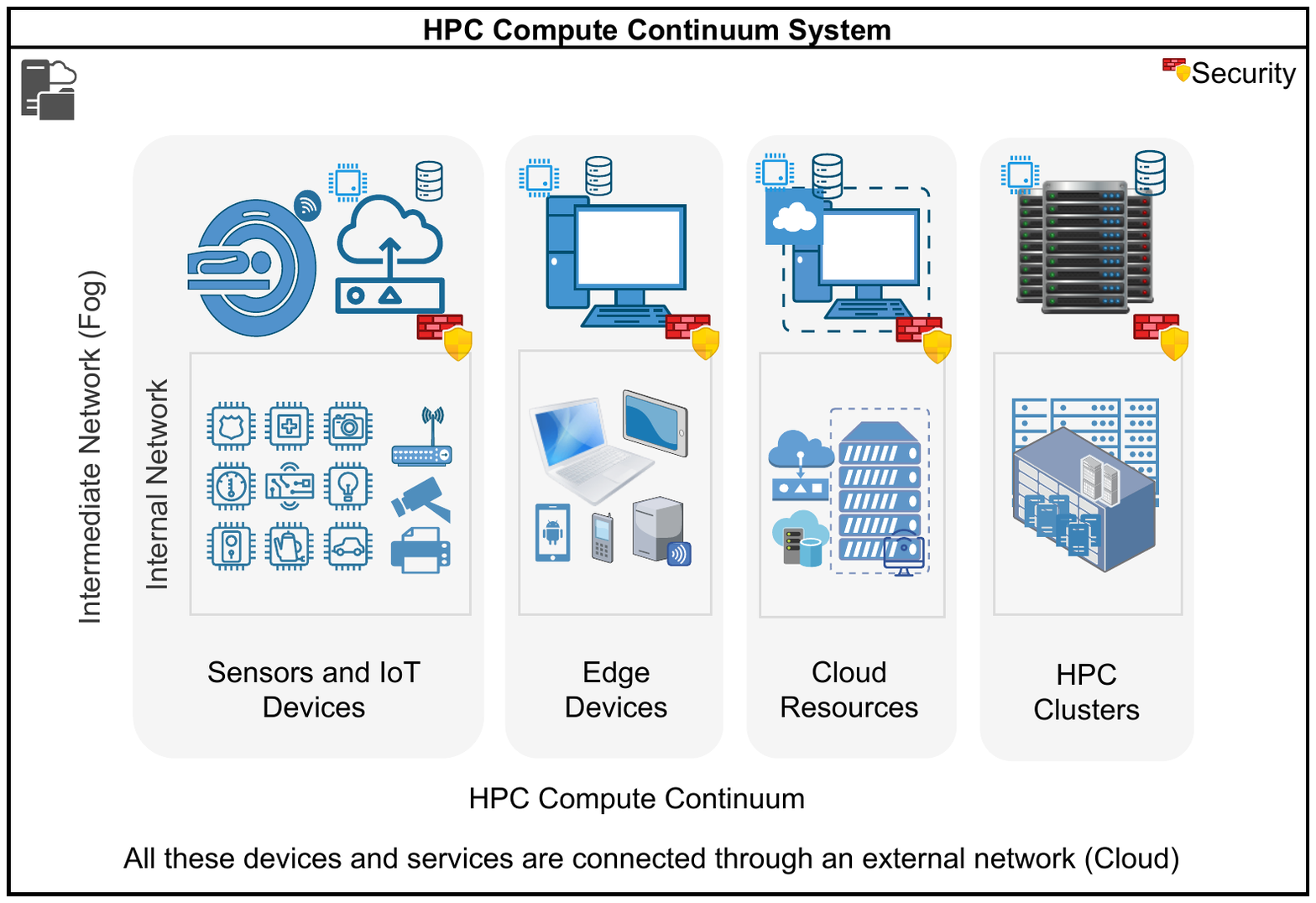}
    \caption{H-HPC-CC (Heterogeneous High-Performance Computing Compute Continuum): Real-world Scenario.}
    \label{fig:HPCComputeContinuum}
    \vspace{-1em}
\end{figure}

High-performance computing (HPC) systems are designed to solve complex problems beyond the scope of conventional computers, often utilizing thousands of parallel processors \cite{nicohabermannDesktopTeraflopExploiting1993, WhatHighPerformanceComputing2022}. Building on von Neumann's foundational model \cite{NeumannArchitecture2022}, modern HPC systems consist of heterogeneous nodes—ranging from IoT devices for data capture and edge nodes for low-latency processing to cloud nodes for scalable storage and traditional HPC nodes for intensive computations. We refer to this integrated system as the \textit{HPC-Compute Continuum} (HPC-CC), as illustrated in \Cref{fig:HPCComputeContinuum}.

\subsection{Workloads: Workflows}
\label{ssec:WorkloadsWorkflows}
\begin{figure}[ht]
    \centering
    \begin{subfigure}[t]{0.45\textwidth}
        \centering
        \includegraphics[width=\textwidth]{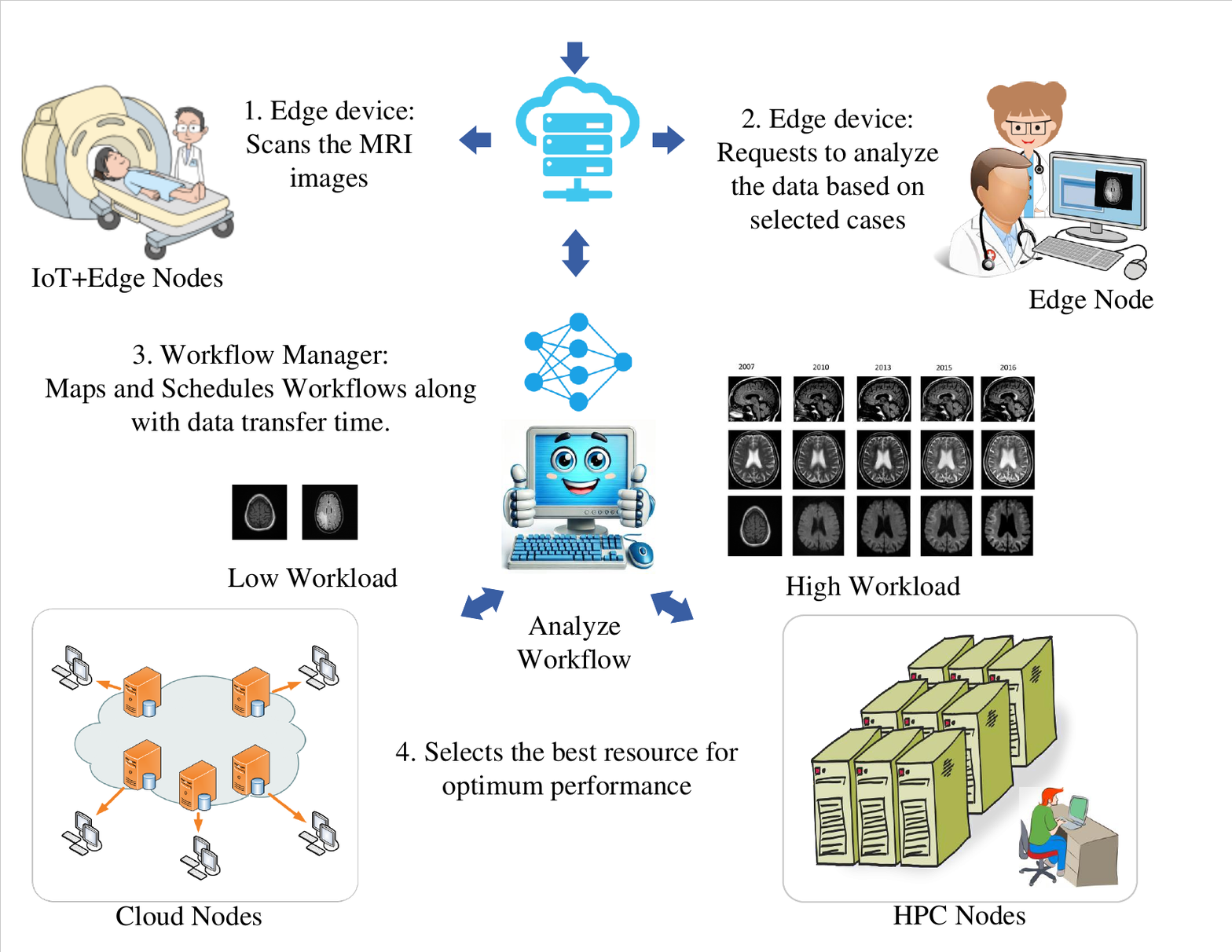}
        \caption{MRI use case with an advance workflow analyzer.}
        \label{fig:MRIUseCase}
    \end{subfigure}\hfill
    \begin{subfigure}[t]{0.45\textwidth}
        \centering
        \includegraphics[width=0.92\linewidth]{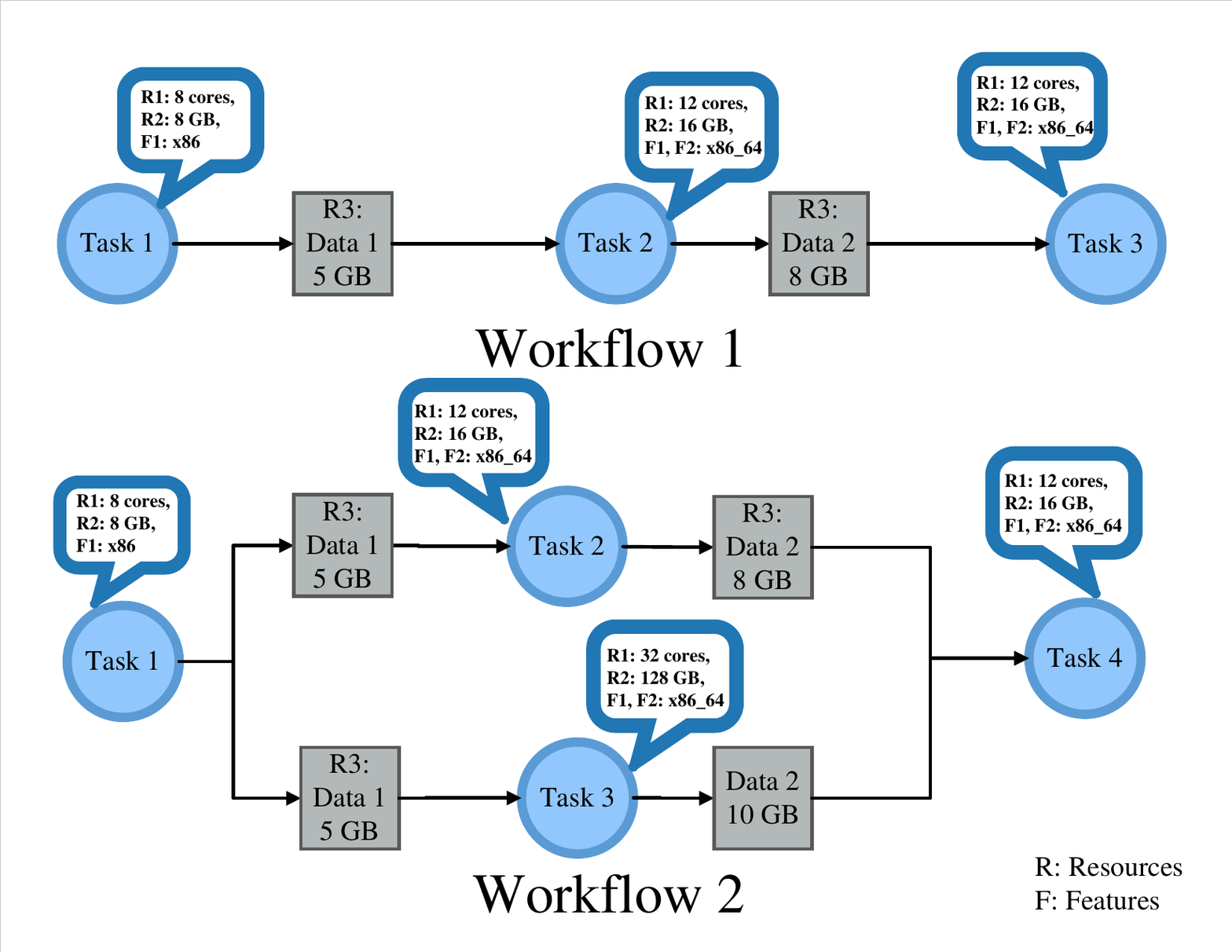}
        \caption{Workflow diagrams for MRI scenarios, showing Petri net workflows with sample compute resource and feature requests along with intermediate data.}
        \label{fig:PetriNetWithResourceAndFeatureRequest}
    \end{subfigure}
    \caption{Diagrams illustrating the MRI use case and corresponding workflows.}
    \label{fig:MRIWorkflow}
    \vspace{-1.5em}
\end{figure}

\begin{figure}[ht]
    \centering
    \includegraphics[width=0.48\textwidth]{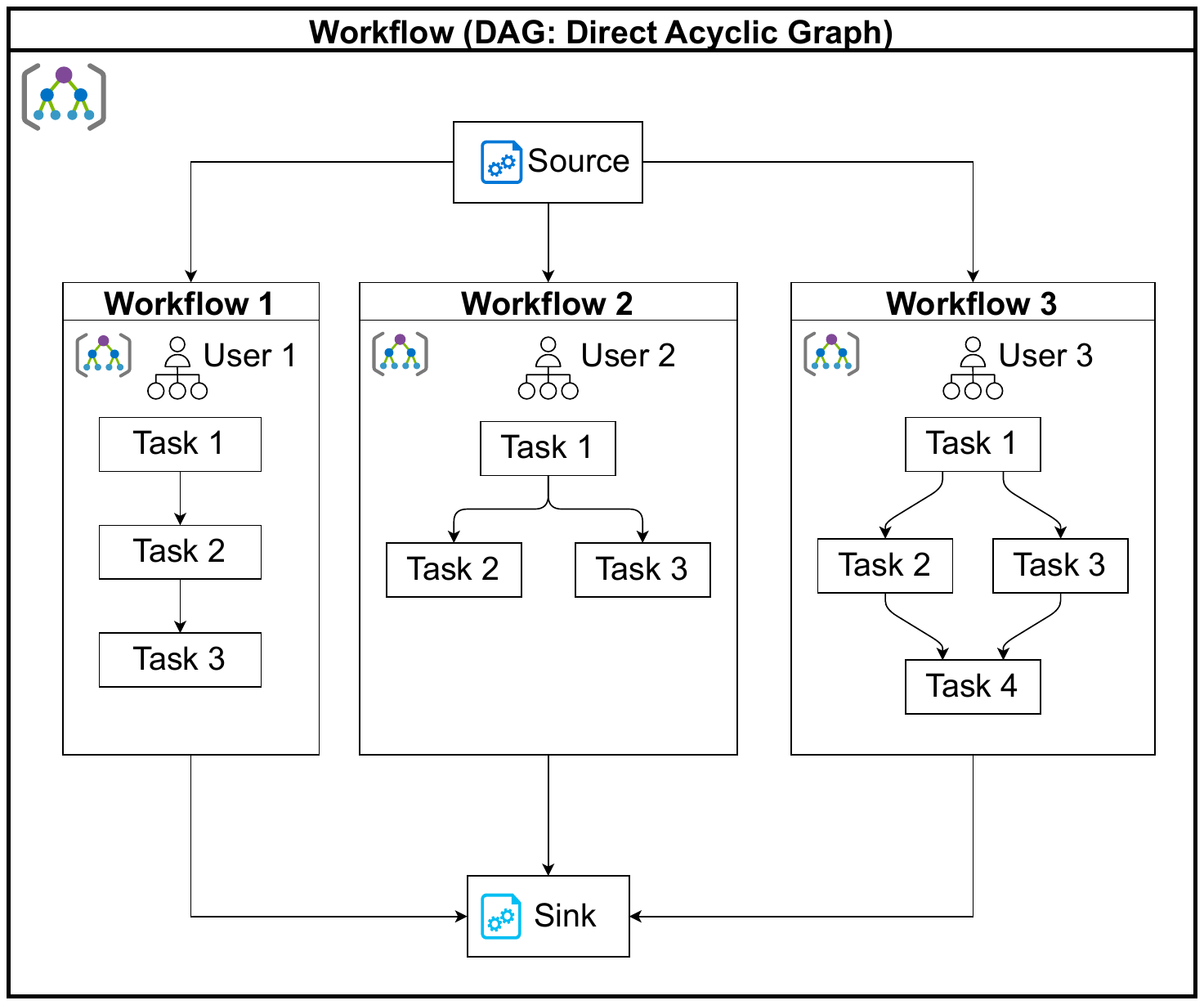} 
    \caption{Heterogeneous workflows on the Compute Continuum for different users.}
    \label{fig:Workflows}
    \vspace{-1em}
\end{figure}

In the HPC-CC, user workloads comprise computational tasks, simulations, and data processing jobs that demand significant resources. For example, consider an MRI (Magnetic Resonance Imaging) use case where an MRI scanner functions as an IoT device to capture high-resolution brain scans. Raw data is initially processed at the edge (e.g., noise reduction and initial reconstruction), then securely stored and further analyzed in the cloud. Advanced image processing tasks are subsequently executed on HPC systems to deliver rapid diagnostic insights. \Cref{fig:MRIWorkflow} illustrates a workflow analyzer and corresponding workflow diagrams for different diagnostic scenarios, while \Cref{fig:Workflows} shows heterogeneous workflows across the compute continuum.

These integrated workflows leverage IoT, edge, cloud, and HPC resources to enhance efficiency and resource utilization. Effective management of such workloads requires sophisticated scheduling and resource management strategies that can handle the diversity and dynamic nature of the underlying infrastructure.

\subsection{Workload Mapping and Scheduling Challenges}
Mapping (allocating appropriate resources) and scheduling (planning task execution) in the HPC-CC aim to minimize makespan and optimize resource usage \cite{fangTaskSchedulingStrategy2022, achterbergConstraintIntegerProgramming2008}. However, challenges such as managing task dependencies (often modeled as DAGs) \cite{alghamdiDynamicClusteringbasedTask2024a, alamResourceawareLoadBalancing2020, agarwalTopologyawareTaskMapping2006} and minimizing data transfer overheads \cite{agarwalDataintensiveScienceTerapixel2011} necessitate integrating additional tools for effective scheduling in complex environments.

\subsection{Available Tools}
We focus on three key tools:
\subsubsection{SLURM \cite{yoo2003slurm}} A widely used workload manager for HPC clusters that efficiently allocates resources based on user-defined configurations.
\subsubsection{Kubernetes \cite{burns2019kubernetes}} A container orchestration platform that excels in dynamic resource allocation for cloud workloads but is less suited for tightly coupled HPC tasks.
\subsubsection{Snakemake \cite{koster2012snakemake}} A workflow management tool that automates reproducible analyses with a rule-based approach. Although Snakemake is robust in managing workflows, it does not include mechanisms for optimal remote resource mapping because it just deploy the workload into remote system's queue.

Integrating these tools allows us to leverage their strengths while mitigating individual limitations. For instance, SLURM's resource allocation can complement Snakemake's workflow management, addressing its lack of remote resource awareness (like a recommendation system).

Our work extends Snakemake's scheduler to incorporate advanced mapping and scheduling constraints derived from our system and workload models. This comprehensive framework forms the basis of our approach, which is further detailed in the up coming sections followed by related work.


\section{Related Work}
\label{sec:RelatedWork}

\subsection{ HPC Systems' Challenges: Reality in Publications}
\label{ssec:GroundRealityPublications}

\subsubsection{Heterogeneous Systems and Load Balancing}
Heterogeneous HPC systems, comprising accelerators like CPUs, GPUs, and FPGAs, leverage specialized unit capabilities but face challenges in load balancing due to architectural differences. Kunzman et al. and Agarwal et al. \cite{KunzmanProgrammingHeterogeneousSystems2011, agarwalDataintensiveScienceTerapixel2011} highlight that disparities in instruction set architectures (ISA) hinder workload distribution, leading to resource underutilization. Dynamic strategies like adaptive scheduling and task offloading \cite{arifApplicationAttunedMemoryManagement2024, alghamdiDynamicClusteringbasedTask2024a} aim to optimize task allocation based on resource performance.

\subsubsection{Scaling and Workload Management in HPC Research}
Managing large-scale workloads across millions of processors remains a critical challenge. Systems like SLURM and PBS (Portable Batch System \cite{openpbs_nodate}) strive to map tasks efficiently but face difficulties with fairness and varied computational demands \cite{paganoMakingControlHigh2024}. Techniques such as gang scheduling and backfilling improve resource utilization, yet balancing fairness and efficiency remains complex \cite{boothNUMAAwareVersionAdaptive2024}.

\subsubsection{Resource Estimation and Communication Overhead}
Accurate resource estimation for dependent tasks poses significant challenges due to communication overhead. Expósito et al. \cite{expositoPerformanceAnalysisHPC2013} propose minimizing communication costs by enhancing data locality and task placement. Emerging approaches, including machine learning-based dynamic adjustments, show promise for optimizing resource allocation but require further development for real-world complexities \cite{tanashImprovingHPCSystem2019, abouelyazidMachineLearningAlgorithms2021}.

\subsubsection{Communication Overhead and System Topologies}
Communication overhead impacts HPC performance significantly, with network topologies (e.g., mesh, torus, fat-tree) playing a critical role. Topology-aware algorithms by Agarwal et al. \cite{agarwalTopologyawareTaskMapping2006} and Li et al. \cite{liAdvancesTopologyAwareScheduling2017} optimize task placement to reduce overhead. Recent advancements in communication-aware scheduling further enhance system efficiency by aligning task placement with network and task communication needs \cite{happeSelfadaptiveHeterogeneousMulticore2013, fangTaskSchedulingStrategy2022}.

\subsection{HPC Systems' Challenges: Reality in Practices}
\label{ssec:GroundRealityPractices}

\subsubsection{Resource Allocation and Usage in Academic Institutions}
In academic HPC environments, shared resources are allocated to departments submitting diverse workloads, ranging from simple scripts to complex simulations. Administrators review detailed user reports to estimate required resources and allocation durations, a task complicated by increasing request volumes \cite{wangRLSchertHPCJob2021, abouelyazidMachineLearningAlgorithms2021}. Dependencies in workflows often cause communication overhead, particularly when tasks are distributed across compute nodes. Efficient utilization requires addressing delays and data transfer overhead to optimize runtime and throughput \cite{singhAutonomicResourceManagement2021}.

\subsubsection{Load Balancing in Heterogeneous Systems - Institutional Experience}
Heterogeneous HPC systems, comprising CPUs, GPUs, and FPGAs, face challenges in balancing workloads among diverse architectures \cite{ApplyComputeProject}. Dynamic resource allocation often struggles to utilize processors efficiently, as highlighted by \cite{alamResourceawareLoadBalancing2020}. CPUs excel in serial tasks, while GPUs are suited for parallel operations, yet poor scheduling can lead to underutilization. Adaptive load balancing strategies offer solutions but remain imperfect \cite{pinchakPracticalHeterogeneousPlaceholder2002}.

\subsubsection{Communication Overhead and Scaling in Real Systems}
As HPC systems scale, inter-node communication becomes a bottleneck. Large-scale installations experience significant runtime increases due to data transfer delays, especially in tightly-coupled tasks \cite{gadbanInvestigatingOverheadREST2020a}. Optimized network topologies, including fat-tree, torus, and mesh, mitigate delays but pose challenges for optimal task placement \cite{kunkelSimulationParallelPrograms2013}. Workload managers must incorporate communication overhead considerations to enhance scheduling efficiency \cite{wangRESCAPEResourceEstimation2024}.

\subsection{Adaptive Workload Management}
Adaptive Workload Management (AWM) is crucial for optimizing resource utilization in High-Performance Computing (HPC) and cloud environments. Traditional static schedulers like SLURM \cite{yoo2003slurm} often struggle to adapt to dynamic workloads. Advanced techniques such as Deep Reinforcement Learning (DRL) \cite{mao2019learning} and feedback-driven auto-scaling \cite{burns2019kubernetes} offer promising solutions. However, balancing competing objectives like energy efficiency, fairness, and scalability remains a significant challenge.

\subsection{Microservices Architecture}
Microservices architecture has emerged as a powerful paradigm for building scalable and resilient distributed systems. By decomposing applications into independent, loosely coupled services, microservices enable efficient resource allocation and fault isolation. Kubernetes \cite{burns2019kubernetes} has become a de facto standard for managing and orchestrating microservices in cloud-native environments. While microservices offer numerous benefits, challenges such as increased operational complexity and communication overhead need to be carefully addressed. Serverless computing \cite{castro2019serverless} is a promising approach to mitigating these challenges by providing a fully managed environment for executing microservices.

\section{Methodology}
\label{sec:ResearchMethodology}
As per the outlined motivation and objectives, we show our employed framework in \Cref{fig:ProcessWorkflow} and also apply respective experimental modeling method to outline a mathematical model for the HPC-CC system and its workload. Based on this experimental modeling we conduct controlled tests to gather data and validate the models through empirical observation and measurement. This approach is particularly useful for validating theoretical concepts, testing hypotheses, and refining model parameters based on real-world observations. We also apply optimization strategies to automate while devising the model. These modeling also helps to estimate the workload completion time along with optimum resource usage in a HPC-CC environment.


\subsection{Methodology for Automated Workload Mapping and Scheduling}
\label{ssec:FrameworkDetails}

\begin{figure*}[ht!]
    \centering
    \includegraphics[width=\linewidth]{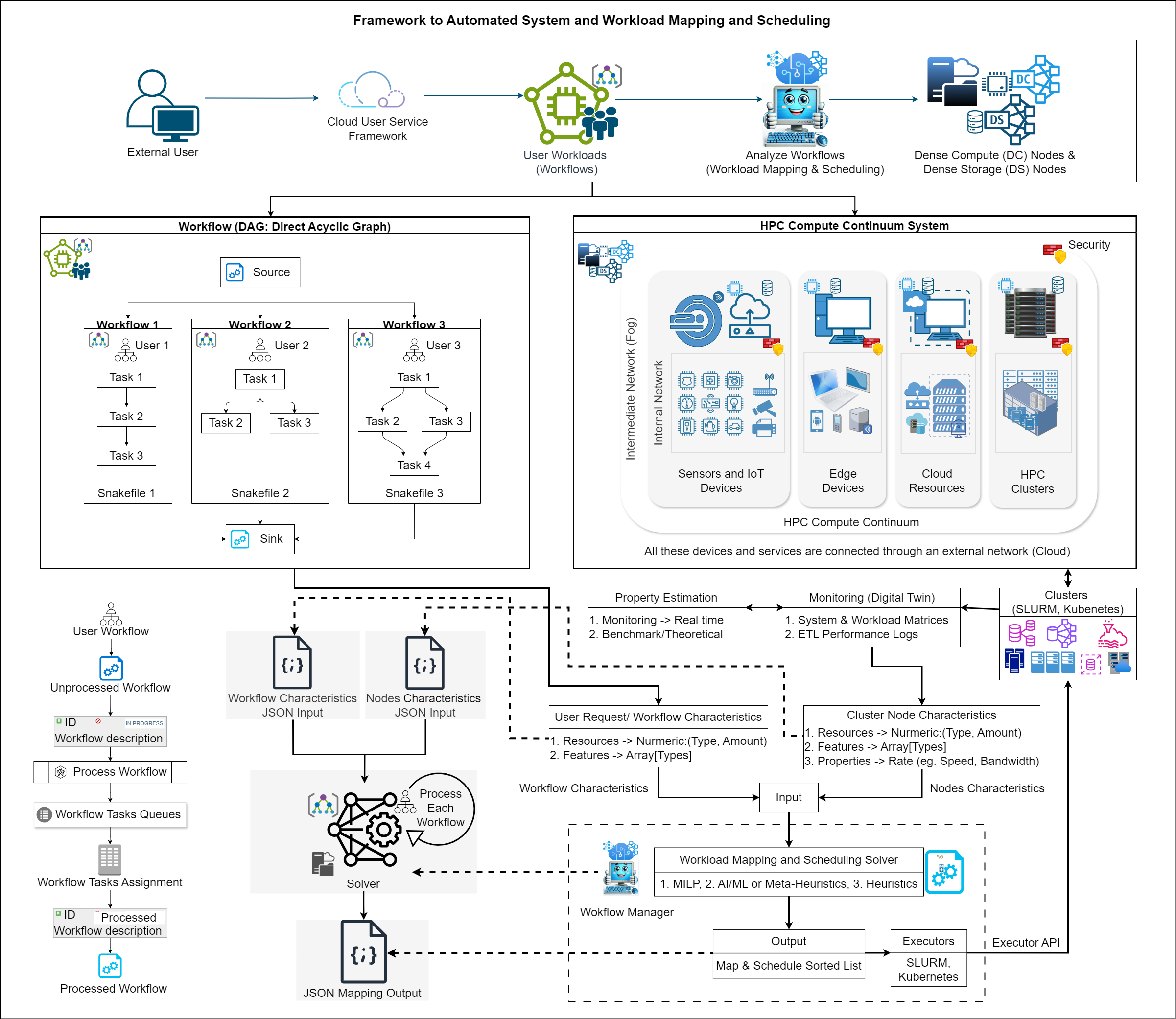}
    \caption{Workload mapping and scheduling framework with auto estimation of incoming workflows.}
    \label{fig:ProcessWorkflow}
\end{figure*}

Figure~\ref{fig:ProcessWorkflow} illustrates the step-by-step process for automating workload mapping and scheduling in an HPC-CC environment. Our methodology integrates a user service framework that allows external users to submit workloads. At regular time intervals—during which the HPC-CC framework checks for new or pending workflows—the following process is executed:

\begin{enumerate}
    \item \textbf{Input Collection:} The first stage involves gathering the input characteristics of the HPC-CC system. Here, the digital twin (monitoring component) provides detailed system and workload performance metrics. In cases where performance data is not available (e.g., the first run), a default seed value based on theoretical estimations is used. All collected system and workload information is formatted in JSON, following the templates presented in \Cref{ssec:InputFormat} (a basic sample).
    
    \item \textbf{Workload Analysis:} The JSON inputs are then supplied to the workload analyzer (solver), which processes the data according to the mathematical models described in \Cref{ssec:SystemAndWorkloadModeling}. Based on the available resources and workflow requirements, the solver maps each workflow to the most optimal system resources, taking into account selected features, constraints, objectives, and resource availability.
    
    \item \textbf{Output Generation:} Once all workflows are processed, the solver produces a sorted list of workflows and their associated tasks in JSON format. This file contains detailed mapping and scheduling information, which is forwarded to the executor.
    
    \item \textbf{Execution and Monitoring:} The executor then dispatches the workflows to the respective SLURM and Kubernetes clusters for execution. After execution, the monitoring component collects logs and performance metrics, updating node properties for subsequent runs.
\end{enumerate}

All of these steps mentioned above are already in testing. The work was divided into different groups under the project framework and each group is responsible for their part. In this study, our primary focus is on the system and workload mapping and scheduling models; therefore, the monitoring component is not discussed in detail.

\subsection{System And Workload Modeling}
\label{ssec:SystemAndWorkloadModeling}

\subsubsection{System Modeling}
\label{sssec:SystemModeling}

This section presents theoretical formulations of system and workload models, linking them to practical applications. Based on \cite{eadlineHighPerformanceComputing2009}, an HPC system consists of clusters $C$, each containing multiple nodes $N$, interconnected via high-speed networks. Nodes comprise resources ($R$), features ($F$), and properties ($P$), forming a heterogeneous or homogeneous configuration.

A node is defined as $N = \{R, F, P\}$:
\begin{itemize}
    \item \textit{Resources ($R$)}: Quantifiable elements like CPU cores, memory, or I/O bandwidth.
    \item \textit{Features ($F$)}: Infrastructure characteristics, such as ISA type or GPU support.
    \item \textit{Properties ($P$)}: Performance metrics, including Processing Speed (PS) or storage throughput.
\end{itemize}

The variables and their detailed definitions are listed in Tables~\ref{tab:ListOfSystemModelVariables}, and \ref{tab:CombinedNodeInfo}.

\begin{table}[H]
    \centering
    \caption{List of system model variables}
    \label{tab:ListOfSystemModelVariables}
    \begin{tabular}{p{1em} p{5em} p{21em}}
    \hline
    \textbf{S.N.} & \textbf{Expression} & \textbf{Definition} \\
    \hline
    1. & $D$ & Data center: Comprises clusters $C$. \\
    2. & $C$ & Cluster: Contains nodes $N$. \\
    3. & $N$ & Node: Defined as $\{R, F, P\}$. \\
    4. & $R$ & Resources: Measurable components of $N$. \\
    5. & $F$ & Features: Node-specific capabilities. \\
    6. & $P$ & Properties: Performance characteristics. \\
    \hline
    \end{tabular}\vspace{-1em}
\end{table}

\subsubsection{Workload Modeling}
\label{sssec:WorkloadModeling}

Workload modeling represents applications as workloads consisting of multiple programs with defined workflows. Each workload requests specific resources and features from system nodes for execution. For example, in SLURM, a \emph{workload} comprises a set of jobs containing one or more tasks with potential dependencies, forming a \emph{workflow}.

As described in \cite{frachtenbergJobSchedulingStrategies2007}, HPC systems aim to maximize resource utilization and minimize idle time. Idle time occurs when the system is operational but not processing workloads due to suboptimal workload sizes or distribution uncertainties. Understanding workloads and system capabilities is crucial for optimizing performance.

Workloads are modeled as a set $L$, as detailed in Table~\ref{tab:ListOfWorkloadModelVariables}. Each workload includes one or more workflows $W$, comprising tasks with dependencies represented by $\delta$. Workflows are modeled as DAGs \cite{vanderaalstWorkflowManagementModels2002}, ensuring no deadlocks or loops.

A DAG is defined as a directed graph with vertices $V$ (tasks) and edges $E$ (dependencies), expressed as $\text{DAG}(\text{Workflow}) = \{V, E\}$, where $E \subseteq \{\{T_{j}, T_{j'}\} \mid T_{j}, T_{j'} \in V \text{ and } T_{j} \neq T_{j'}\}$, with $T_{j'} \rightarrow T_{j}$ denoting a dependency. \Cref{fig:PetriNetWithResourceAndFeatureRequest}, \Cref{fig:Workflows} and \Cref{fig:STGSWorkflows} illustrate DAGs and workflows \cite{vanderaalstWorkflowManagementModels2002}. \Cref{fig:PetriNetWithResourceAndFeatureRequest} also includes a Petri-net representation of workflows, where transitions correspond to tasks $T$ and places to data $D_d$, highlighting relationships between tasks and data flows. Each task $T$ in a workflow is represented as a tuple $T = \{R, F, U, \delta\}$:\newline - $R$: Requested resources ($R^r_T$),
\newline - $F$: Required features ($F^f_T$),
\newline - $U$: Resource usage ($U^r_T$),
\newline - $\delta$: Dependencies between tasks $j$ and $j'$. \newline Dependencies such as data transfers are analyzed to estimate transfer times between nodes. Resource usage for each task is evaluated using the performance characteristics in \Cref{tab:CombinedNodeInfo}.

\begin{table}
    \centering
    \caption{List of workload model variables}
    \label{tab:ListOfWorkloadModelVariables}
    \begin{tabular}{p{1em} p{3.5em} p{21em}}
    \hline
    \textbf{S.N.} & Expression & Definition \\
    \hline
    1. & $L$ & $Workload$: A set of workflows $\{W_1, W_2, ..., W_w\}$, where $w$ is the number of workflows. \\
    2. & $W$ & $Workflow$: A set of tasks $W = (\{T_1, T_2, ..., T_{|T|}\}, s)$, with $|T|$ tasks and $s$ as submission time. \\
    3. & $T$ & A $Task$ represented as $T=\{R, F, U, \delta\}$. \\
    4. & $R$ & Requested resources $R:R^r_{T}$, as defined in \Cref{tab:CombinedNodeInfo}. \\
    5. & $F$ & Requested features $F:F^f_{T}$, as defined in \Cref{tab:CombinedNodeInfo}. \\
    6. & $U$ & Resource usage $U:U^r_T$, calculated based on node resource usage. Further explained in \Cref{sssec:ResourceUsageModel}. \\
    7. & $\delta$ & Dependencies $\delta(j, j', ..., |T|)$, showing relationships between task instances in a workflow. \\
    \hline
    \end{tabular}
\end{table}


\begin{table}[ht]
    \centering
    \caption{Node Resources, Features, and Properties}
    \label{tab:CombinedNodeInfo}
    \begin{tabular}{p{1em} p{5em} p{20em}}
    \hline
    \textbf{S.N.} & \textbf{Expression} & \textbf{Definition} \\
    \hline
    \multicolumn{3}{l}{\textbf{Node Resources}} \\
    1. & $R^1$ & Number of cores. \\
    2. & $R^2$ & Memory capacity (GB). \\
    3. & $R^3$ & Storage capacity (GB). \\
    \hline
    \multicolumn{3}{l}{\textbf{Node Features}} \\
    4. & $F^1$ & Processor ISA: x86 (CPU). \\
    5. & $F^2$ & Processor ISA: x64 (GPU). \\
    6. & $F^3$ & Memory type: DDR4. \\
    7. & $F^4$ & Memory type: DDR5. \\
    8. & $F^5$ & Storage type: HDD. \\
    9. & $F^6$ & Storage type: SSD. \\
    10. & $F^7$ & Network: Omni-Path. \\
    11. & $F^8$ & Network: InfiniBand. \\
    \hline
    \multicolumn{3}{l}{\textbf{Node Properties}} \\
    12. & $P^1$ & Processing Speed (e.g., Processor base clock). \\
    13. & $P^2$ & Processing Speed (e.g., FLOP/s). \\
    14. & $P^3$ & Data Transfer Rate (e.g., PCI/Bus). \\
    \hline
    \end{tabular}
    \vspace{-1em}
\end{table}

\subsection{Workload Mapping and Scheduling Characteristics}
\label{ssec:WorkloadMappingAndSchedulingCharacteristics}

The defined system and workload models enable analysis of workload mapping and scheduling within heterogeneous HPC environments. Tasks in the queue request specific resources and features, posing the challenge of meeting these requirements efficiently.

\emph{Mapping} optimally allocates node resources to tasks, while \emph{scheduling} aims to minimize resource inefficiencies and overall completion time. Optimizing workload mapping and scheduling addresses challenges like overutilization and underutilization, enhancing performance.

\subsubsection{System and Workload Relation}
\label{sssec:SystemAndWorkloadRelation}

Nodes and workloads are represented as tuples of resources, features, and performance characteristics. Tasks request specific resources and features that must be provided by nodes possessing the required attributes. This relationship is defined in \Cref{equ:ResourceAndFeatureRelation}.

\begin{equation}
    \label{equ:ResourceAndFeatureRelation}
    R^r_T \subseteq R^r_N \quad \text{and} \quad F^f_T \subseteq F^f_N
\end{equation}

\subsubsection{Resource Allocation (Relates to Mapping)}
\label{sssec:ResourceAllocationModel}

Let $i$ represent a node instance $N$ in cluster $C$, and $j$ a task in workflow $W$. Task $j$ requests resources $R^r_j$ and is assigned to node $i$ if $i$ has sufficient available resources $R^r_i$. A binary variable $x_{ij}$ maps task $j$ to node $i$, where $x_{ij} = R^r_j / R^r_i$. Resource allocation is valid if $x_{ij} \leq 1$, as expressed in \Cref{equ:ResourceCapacity}:

\small
\begin{equation}
    \label{equ:ResourceCapacity}    
    x_{ij}=\frac{R^r_{j}}{R^r_{i}}   
    \begin{cases} 
        x_{ij} \leq 1 & \text{allowed;}  \\
        x_{ij} > 1 & \text{not allowed.} 
    \end{cases}      
\end{equation}
\normalsize

\textit{Note:} For dynamic allocations or multiple tasks per node, $x_{ij}$ must be computed cumulatively.

\subsubsection{Resource Usage (Relates to Mapping and Scheduling)}
\label{sssec:ResourceUsageModel}

The resource usage $U_{ij}$ for task $j$ on node $i$ is calculated using:

\begin{equation}
    \label{equ:ResourceUsage}
    U_{ij} = R_j \cdot \left( \frac{R_i}{\sum_{i'} R_{i'}} \right)
\end{equation}

Here, $R_j$ denotes the resources requested by task $j$, and $\sum_{i'} R_{i'}$ is the total available resources across nodes. In fixed-resource scenarios, $U_j = R_j$. For heterogeneous nodes, \Cref{equ:ResourceUsage} adjusts for capacity differences.

\subsubsection{Resource Duration (Relates to Scheduling)}
\label{sssec:ResourceDuration}

The duration $d_{ij}$ for task $j$ on node $i$, based on requested resources $R_j$ and node performance $P^p_i$, is given by:

\begin{equation}
    \label{equ:ResourceDuration}
    d_{ij} = \frac{R_j}{P^p_i}
\end{equation}

In workflows with dependencies, inter-node data transfers affect duration. Transfer time $d^3_{t:ii'j}$ for task $j'$ between nodes $i$ and $i'$ is:

\begin{equation}
    \label{equ:CalculateDataTransferTime}
    d_{t:ii'j} = \frac{R^3_j}{P^3_{ii'}}
\end{equation}

Here, $P^3_{ii'}$ is the Data Transfer Rate (DTR) between nodes $i$ and $i'$ for data $R^3_j$.

\subsubsection{Total Resource Usage and Duration}
\label{sssec:TotalResourceDuration}

The total duration $d_{ij}$ for task $j$ on node $i$, including resource usage and data transfer, is:

\begin{equation}
    \label{equ:ResourceUsageDuration}
    d_{ij} = \max_{r \in |R|} \left( d^r_{ij} + d^r_{t:ii'j} \right)
\end{equation}

This accounts for the maximum duration across all requested resources and associated data transfer times.

\subsubsection{Multi-Objective Function}
\label{ssec:ObjectiveFunction}

The task and node numbers are constrained to natural numbers, excluding negative integers and non-integers. The derived models are linear, enabling the objective function to be defined using linear programming principles \cite{LinearProgrammingExtensions1998}:

\begin{equation}
    \label{equ:LqCostFunction}
    \text{minimize} \quad f(x) = \sum_{i=1}^n c_i x_i
\end{equation}

Here, $f(x)$ represents the total cost, $x_i$ denotes decision variables, $c_i$ are cost coefficients, and $n$ is the total number of variables. The function adheres to linear constraints, including:
\begin{enumerate}[(a)]
    \item Simple bounds, e.g., $x_i \leq 0$.
    \item Equality constraints, e.g., $\sum_{i=1}^n a_i x_i = b_i$.
    \item Inequality constraints, e.g., $\sum_{i=1}^n a_i x_i \leq b_i$.
\end{enumerate}

The multi-objective function, aiming to optimize resource usage and minimize makespan, is expressed as:

\begin{equation}
    \label{equ:MLObjectiveFunction}
    \text{Minimize} \quad \alpha \sum_{j \in |T|} \sum_{i \in |N|} U_{ij} x_{ij} + \beta C_{\text{max}}
\end{equation}

Here, $x_{ij}$ is a binary decision variable:
\[
x_{ij} = 
\begin{cases} 
1 & \text{if node } i \text{ is assigned to task } j, \\
0 & \text{otherwise.}
\end{cases}
\]

The finish time $f_j$ is:
\[
f_j = s_j + d_{ij} x_{ij} \quad \forall j \in |T|, \forall i \in |N|
\]

Data transfer constraints are defined by:
\[
s_j \geq f_{j'} + d^r_{j'} + d^r_{t:ii'j'} \quad \forall (j', j) \in \delta, \forall i (i \neq i') \in |N|.
\]

\subsubsection{Introducing Applied Constraints}
\label{sssec:Constraints}

Key constraints include:
\begin{enumerate}[a)]
    \item \textbf{Task Assignment:} Each task is assigned to exactly one node:
    \begin{equation}
        \label{equ:CstTaskAssignment}
        \sum_{i=1}^{|N|} x_{ij} = 1 \quad \forall j \in |T|.
    \end{equation}

    \item \textbf{Resource Constraints:} Tasks assigned to a node cannot exceed its capacity:
    \begin{equation}
        \label{equ:CstNodeCapacity}
        \sum_{j=1}^{|T|} x_{ij} \leq R_i \quad \forall i \in |N|.
    \end{equation}

    \item \textbf{Feature Constraints:} Tasks must be assigned to nodes meeting required features:
    \begin{equation}
        \label{equ:CstTaskFeature}
        \sum_{j=1}^{|T|} x_{ij} = 1_{F^f_j \subseteq F^f_i} \quad \forall i \in |N|.
    \end{equation}

    \item \textbf{Task Dependency with Data Migration:} Dependencies ensure:
    \begin{equation}
        \label{equ:CstTaskDependency}
        s_j \geq f_{j'} + d_{ij'} \quad \forall (j', j) \in \delta, \forall i \in |N|.
    \end{equation}
    Additionally:
7    \begin{equation}
        \label{equ:CstTaskDependencyBinary}
        y_{ii'j} \geq x_{ij} + x_{i'j'} - 1,
    \end{equation}
    ensuring $y_{ii'j} = 1$ if tasks are assigned to different nodes.
\end{enumerate}

This conclude the modeling part. Now we will validate the model in the next section.

\section{Validation and Illustration}
\label{sec:ObservationAndDiscussion}
To illustrate our system and workload modeling, we first present a small-scale example that spans from workload specification to the actual job schedule creation. This example enables the reader to follow our approach from the workload definition to the resulted schedule  observation.

We begin by showing how a standard Snakemake workflow (\Cref{fig:OriginalSnakefile}) can be used to generate a workload model. Typically, users must manually specify the target execution node for each rule—hardwiring the node assignments directly into the Snakefile. This approach is suboptimal and inflexible, in case of remote execution. In contrast, our system eliminates the need for such manual specifications.

Furthermore, when users provide a system specification of the compute continuum. For instance, take the MRI scenario (\Cref{tab:SampleWorkloadCharacteristics}) with three node types, as detailed in \Cref{tab:SampleNodeCharacteristics}. Internally, this specification is specified in a JSON format (e.g., \Cref{fig:InputNodes}). The scheduler uses this input to capture node characteristics and combines them with workload attributes derived from a customized Snakefile that includes additional custom attributes (e.g., \Cref{fig:CustomSnakefile}).

The JSON input formats for both system (\Cref{fig:InputNodes}) and workflow characteristics (\Cref{fig:InputWorkflow}) adheres to the standard Snakemake configuration approach.

\subsection{User-Specified Snakemake Workflow} 
\label{ssec:InputValidation} 

Our JSON input formats (shown in \Cref{fig:CustomSnakefile}) align with best practices for Snakemake configuration. In Snakemake, system and workflow parameters are typically stored in a JSON (or YAML) file and loaded into the Snakefile via the \texttt{configfile} directive, as illustrated in \Cref{fig:OriginalSnakefile} \cite{Koster:2012}. This design facilitates a direct mapping of node properties (e.g., \texttt{cores}, \texttt{memory}, with auxiliary attributes like \texttt{features}) and task requirements (e.g., \texttt{memory\_required}, \texttt{dependencies}) to resource allocation rules for enhancements. This helps to address the challenges related to remote execution. We understand, it is a big challenge to completely address the user and host requirements but this adds an option to extend when able to properly collect the remote system and the workload characteristics.

\begin{figure}[ht!]
    \small
    \vspace{-1em}
    \begin{lstlisting}[language=bash, morekeywords={input,output,resources, run, rule,slurm_partition}]
rule T1:
    input:
        experiment.conf
    output:
        product1.dat
    resources:
        mem_mb   = 1024
        slurm_partition = gpuNode
        runtime = 10:00:00
    run:
        # Execute shell command/script

rule T2:
    input:
        product1.dat
    output:
        product2.dat
    resources:
        slurm_partition = cpuNode
    ...
    \end{lstlisting}
    \caption{Original Snakefile rule.}
    \label{fig:OriginalSnakefile}
    \vspace{-2em}
\end{figure}

\begin{figure}[ht!]
    \small
    \vspace{-1em}
    \begin{lstlisting}[language=bash, morekeywords={input,output,resources, run, rule}]
rule T1:                        # dependencies
    input:
        experiment.conf
    output:
        product1.dat,           
    resources:
        mem_mb = [1024]         # memory_required, (R2)
        features = ["F1", "F2"]       # requested features
        data = 2GiB             # estimated output size, (R3)
        duration = [1000]       # usage, must specify all in seconds, (dij)
    run:
        # Execute shell command/script
rule T2:
    input:
        product1.dat
    output:
        product2.dat
    resources:
        features = ["F1"]
        ...
    \end{lstlisting}
    \caption{Example Snakefile rule (annotating the workflow).}
    \label{fig:CustomSnakefile}
\end{figure}

\subsection{System and Workload Input Format}
\label{ssec:InputFormat}


\begin{figure}[H]
    \small
    \begin{verbatim}
{"nodes": {
    "Node1": {
        "cores": [4],
        "memory": [1024],
        "features": ["F1"],
        "processing_speed": [1024],
        "data_transfer_rate": [100]
    },
    "Node2": {
        "cores": 12,
        ...
    }
}}
    \end{verbatim}
    \caption{Example system characteristics in JSON.}
    \label{fig:InputNodes}
\end{figure}

\begin{figure}[H]
    \vspace{-1em}
    \small
    \begin{verbatim}
{"Workflow 1": {
    "tasks": {
        "T1": {
            "cores": [4],
            "memory_required": [1024],
            "features": ["F1"],
            "data": 1024,
            "duration": [10],
            "dependencies": []
        }, ...
    }}}
    \end{verbatim}
    \caption{Example workflow in JSON.}
    \label{fig:InputWorkflow}
\end{figure}

\subsection{Snakemake Scheduler and its Algorithms}
\label{ssec:SnakemakeScheduler}
Further, Snakemake’s scheduler \cite{Koster:2012} constructs a directed acyclic graph (DAG) from user-defined rules and dependencies. It applies topological sorting using ILP (Integer Linear Programming) with greedy strategies to dispatch jobs as quick as resource constraints (e.g., cores, memory) are met, also it does lazy evaluation to minimize complexity. 

We could extend the Snakemake Scheduler class to incorporate the mapping and scheduling models, with respective constraints and objectives applying different optimization techniques like (MILP, meta-heuristic, or heuristic detailed in Table~\ref{tab:ToolsAndTechniquesList}) as well as incorporating the time threshold strategies as in Snakemake. Onwards, based on the accumulated optimum mapping and scheduling output information we could deploy the respective workflows to the remote systems using Snakemake's remote executor plugins or the customizable APIs.

\begin{figure}
    \centering
    \includegraphics[width=\linewidth]{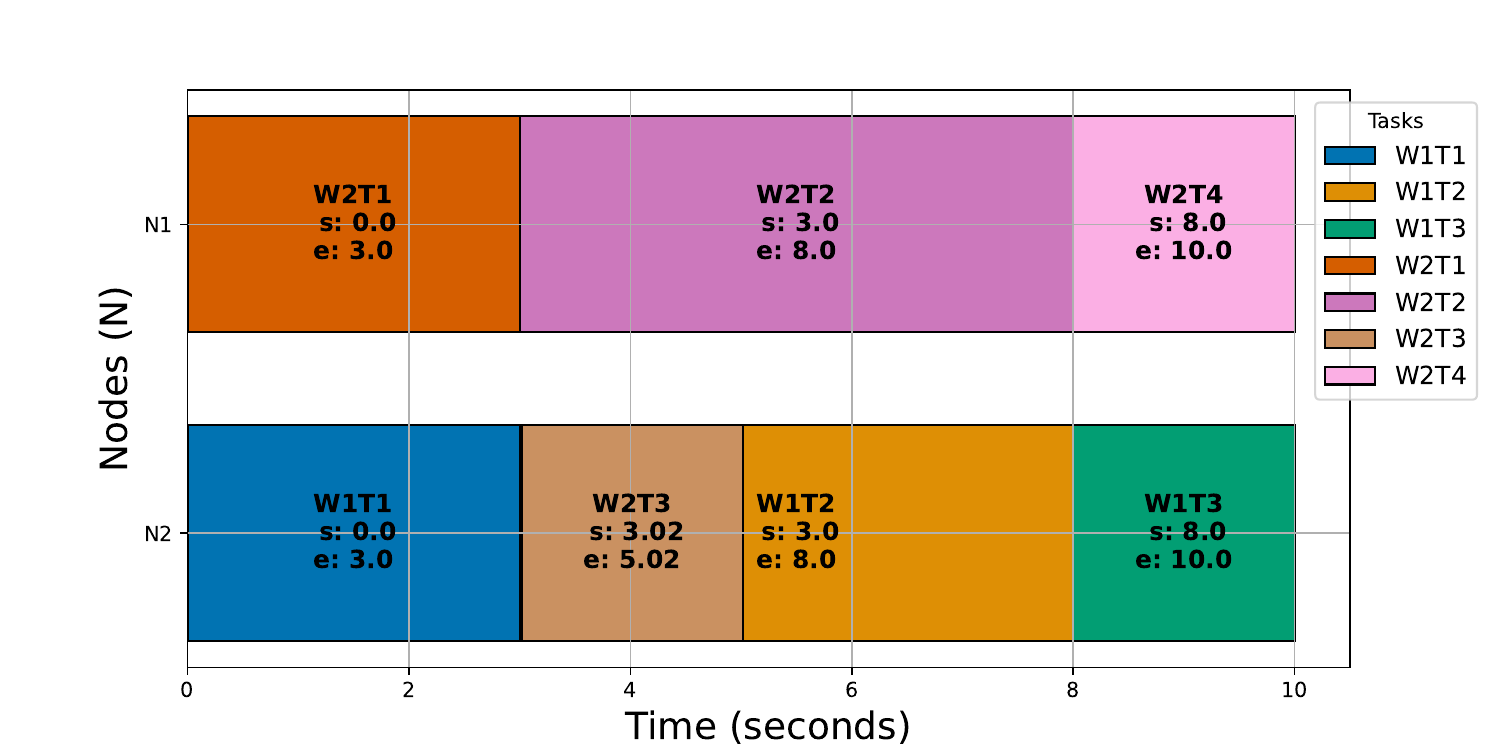}
    \caption{MILP: PuLP obtained optimum feasible solution (matches the manual estimation).}
    \label{fig:MIP_OptimumScheduleGanttChart}
    \vspace{-1em}
\end{figure}

Besides, to validate the extended framework, we test the respective solver with the MRI use case (as shown in Section~\ref{sec:Background}) taking arbitrary system and workload characteristics provided in Tables~\ref{tab:SampleNodeCharacteristics} and \ref{tab:SampleWorkloadCharacteristics}. The manual optimal solution (Table~\ref{tab:SolutionMIP}) evaluated apply the algorithm (\Cref{alg:ResourceAllocationAndSchedulingWithDataTransferTime}) closely matches with the programmatically generated MILP outputs, as demonstrated in Figure~\ref{fig:MIP_OptimumScheduleGanttChart}.

\begin{table}[ht]
    \centering
    \caption{Sample node characteristics.}
    \label{tab:SampleNodeCharacteristics}   
    \begin{tabular}{p{1em}p{10.5em}llp{6em}}
        \hline
        \textbf{S.N.} & \textbf{Nodes} & \textbf{$N_1$} & \textbf{$N_2$}  & \textbf{$N_3$...} \\
        \hline
        1. & Node Type & Node 1 & Node 2 & Node 3 \\ 
        2. & Core ($R^1$) & 8 & 48 & 2572 \\         
        3. & Storage/node ($R^3$ in TB) & 0.5 & 20 & 210 \\  
        4. & Features & $F^1$ & $F^1$, $F^2$ & $F^1$, $F^2$, $F^3$  \\ 
        5. & DTR ($P^3$ in GB/s) & 100 & 100 & 100 \\        
        6. & PS ($P^2$ FLOPs) & 1 & 1 & 1  \\ 
        \hline
    \end{tabular}
    \vspace{-1em}
\end{table}

\begin{table}[ht]
    \centering
    \caption{Sample workload characteristics from \Cref{fig:PetriNetWithResourceAndFeatureRequest}.}
    \label{tab:SampleWorkloadCharacteristics}
    \begin{tabular}{p{1em}p{1em}p{2em}p{2em}p{2em}|p{4.5em}p{3.5em}|p{3em}}
        \hline
        \textbf{$W$} & \textbf{$T$} & \textbf{($R^1$)} & \textbf{($F^f$)} & \textbf{($R^3$)} & \textbf{($d^r_{ij}$)} & \textbf{($d^r_{t:ii'j'}$)} & \textbf{$\delta(j,j')$} \\
        Id & Id & & & & $N1, N2, N3$ & (same) & \\
        \hline
              & $T_1$ & 8  & $F_1$      & 2  & (3, 3, 3) & 0.02 & - \\ 
        $W_1$ & $T_2$ & 12 & $F_1, F_2$ & 5  & (5, 5, 5) & 0.05 & $\delta(T_1,T_2)$ \\ 
              & $T_3$ & 12 & $F_1, F_2$ & 8  & (2, 2, 2) & 0.08 & $\delta(T_2,T_3)$ \\ 
        \hline
              & $T_1$ & 8  & $F_1$      & 2  & (3, 3, 3) & 0.02 & - \\ 
        $W_2$ & $T_2$ & 12 & $F_1, F_2$ & 5  & (5, 5, 5) & 0.05 & $\delta(T_1,T_2)$ \\ 
              & $T_3$ & 32 & $F_1, F_2$ & 5  & (2, 2, 2) & 0.05 & $\delta(T_1,T_3)$ \\ 
              & $T_4$ & 12 & $F_1, F_2$ & 10 & (2, 2, 2) & 0.10 & $\delta(T_2,T_4)$, $\delta(T_3,T_4)$ \\ 
        \hline
    \end{tabular}
    \vspace{-1em}
\end{table}

\begin{table}[ht]
    \centering

    \caption{Manually estimated feasible, optimum solution.}
    \label{tab:SolutionMIP}
    \begin{tabular}{p{2.5em}p{3em}p{1.5em}p{2.5em}p{1.5em}p{1.5em}p{3em}p{3em}}
        \hline\small
        \textbf{Status} & \textbf{Workflow} & \textbf{Task} & \textbf{Optimal Node} & \textbf{Start in\,s} & \textbf{End in\,s} & \textbf{Resource Usage} & \textbf{Makespan in\,s} \\
        \hline
                      &                   & $T_1$ & $N_2$ & 0.0 & 3.0 &  & \\
        \textbf{Optimal} & $W_1$ & $T_2$ & $N_2$ & 3.0 & 8.0 &  & \\
                      &                   & $T_3$ & $N_2$ & 8.0 & 10.0 &  & \\
        \hline
                      & \textbf{Total}    &       &      &      &      & 32.0 & 10.0 \\
        \hline
                      &                   & $T_1$ & $N_1$ & 0.0 & 3.0 &  & \\
        \textbf{Optimal} & $W_2$ & $T_2$ & $N_1$ & 3.0 & 8.0 &  & \\
                      &                   & $T_3$ & $N_2$ & 3.02 & 5.02 &  & \\
                      &                   & $T_4$ & $N_1$ & 8.0 & 10.0 &  & \\
        \hline
                      & \textbf{Total}    &       &      &      &      & 64.0 & 10.0 \\
        \hline
    \end{tabular}
    \vspace{-1em}
\end{table}

\begin{algorithm}
\caption{Resource Allocation and Scheduling with Data Transfer Time}
\begin{algorithmic}[1]

\STATE \textbf{Input:} 
\STATE \hspace{1em} Nodes: $\{(R_i, F_i)\}_{i \in |N|}$ \hfill \# Resources and features of each node
\STATE \hspace{1em} Tasks: $\{(R_j, F_j, d_j, \delta_j)\}_{j \in |T|}$ \hfill \# Resources, features, duration, and dependencies of each task
\STATE \hspace{1em} Data Transfer Time: $\{d_{jj'}\}_{(j,j') \in \delta}$ \hfill \# Data transfer times between tasks

\STATE \textbf{Variables:} 
\STATE \hspace{1em} $x_{ij} \in \{0, 1\}$ \hfill \# Binary variable indicating if task $j$ is assigned to node $i$
\STATE \hspace{1em} $s_j \in \mathbb{R}^+$ \hfill \# Start time of task $j$
\STATE \hspace{1em} $f_j \in \mathbb{R}^+$ \hfill \# Finish time of task $j$
\STATE \hspace{1em} $C_{\text{max}} \in \mathbb{R}^+$ \hfill \# Maximum finish time of all tasks
\STATE \hspace{1em} $y_{jj'} \in \{0, 1\}$ \hfill \# Binary variable for data transfer constraint

\STATE \textbf{Objective:}
\STATE \hspace{1em} Minimize: $ \alpha \cdot \sum_{j \in |T|} \sum_{i \in |N|} U_j \cdot x_{ij} + \beta \cdot  C_{\text{max}}$

\STATE \textbf{Constraints:}
\STATE \hspace{1em} \textbf{Assignment Constraints:}
\FOR{each task $j \in |T|$}
    \STATE $\sum_{i \in |N|} x_{ij} = 1$ \hfill \# Each task must be assigned to exactly one node
\ENDFOR

\STATE \hspace{1em} \textbf{Resource Constraints:}
\FOR{each node $i \in |N|$}
    \STATE $\sum_{j \in |T|} U_j \cdot x_{ij} \leq R_i$ \hfill \# Node resource capacity
\ENDFOR

\STATE \hspace{1em} \textbf{Feature Constraints:}
\FOR{each node $i \in |N|$}
    \STATE $\sum_{j \in |T|} x_{ij} = 1_{F^f_j \subseteq F^f_i}$ \hfill \# Node with requested features
\ENDFOR

\STATE \hspace{1em} \textbf{Timing Constraints:}
\FOR{each task $j \in |T|$}
    \STATE $f_j = s_j + d_j$ \hfill \# Finish time is the start time plus the duration of the task
\ENDFOR

\STATE \hspace{1em} \textbf{Makespan Constraint:}
\FOR{each task $j \in |T|$}
    \STATE $C_{\text{max}} \geq f_j$ \hfill \# Makespan is the maximum finish time of all tasks
\ENDFOR

\STATE \hspace{1em} \textbf{Dependency Constraints:}
\FOR{each pair of tasks $(j, j') \in \delta$}
    \STATE $s_{j'} \geq f_j + d_{jj'} \cdot (1 - y_{jj'})$ \hfill \# Task $j'$ starts after $j$ finishes, considering data transfer time
    \STATE \textbf{Ensure Data Transfer Variable:}
    \STATE $y_{jj'} \geq x_{ij} + x_{i'j'} - 1$ \hfill \# Ensure $y_{jj'}$ is 1 if tasks are assigned to different nodes
\ENDFOR

\STATE \textbf{Solve the MILP:}
\STATE Use a MILP solver to solve the problem

\STATE \textbf{Output:} Optimal task assignments, start times, finish times, and makespan

\end{algorithmic}
\label{alg:ResourceAllocationAndSchedulingWithDataTransferTime}
\end{algorithm}

\begin{table}[ht]
    \centering    
    \caption{List of tools and techniques evaluated.}
    \label{tab:ToolsAndTechniquesList}
    \begin{tabular}{p{7em}|p{21em}}
        \hline 
        \textbf{Optimum Solution: MILP Tools} & \textbf{Approximate Solution: Meta-Heuristics (MH) \& Heuristics (H)} \\
        \hline 
        \begin{tabular}{p{5.5em}} 
            \\
            \\                   
            PuLP \cite{OptimizationPuLPPuLP} \\
            SCIP \cite{bestuzhevaEnablingResearchSCIP2023} \\
            Gurobi \cite{GurobiOptimizer} \\ 
            OR-Tools \cite{ORTools}  \\  
            \\
            \\
            \hline
        \end{tabular}
        & 
        \begin{tabular}{p{20em}}
            MH: Nature Inspired \\
            \hline
            GA: Genetic Algorithm \cite{fatehiEnergyAwareMulti2021} \\
            PSO: Particle Swarm Optimization \cite{gadParticleSwarmOptimization2022} \\
            ACO: Ant Colony Optimization \cite{jiangEnergyefficientSchedulingFlexible2022} \\
            SA: Simulated Annealing \cite{delahayeSimulatedAnnealingBasics2019} \\   
            \hline
            H: Sorting Techniques \\
            \hline
            HEFT: Heterogeneous Earliest Finish Time \cite{maOptimizedWorkflowScheduling2021} \\
            OLB: Opportunistic Load Balancing \cite{mishraLoadBalancingCloud2020} \\ 
            \hline
        \end{tabular}
        \vspace{-0.4em}
    \end{tabular}
    \vspace{-1em}
\end{table}

\begin{figure*}[ht]
    \centering
    \small
    \begin{subfigure}[t]{0.32\textwidth}
        \raggedright
        \includegraphics[width=\textwidth]{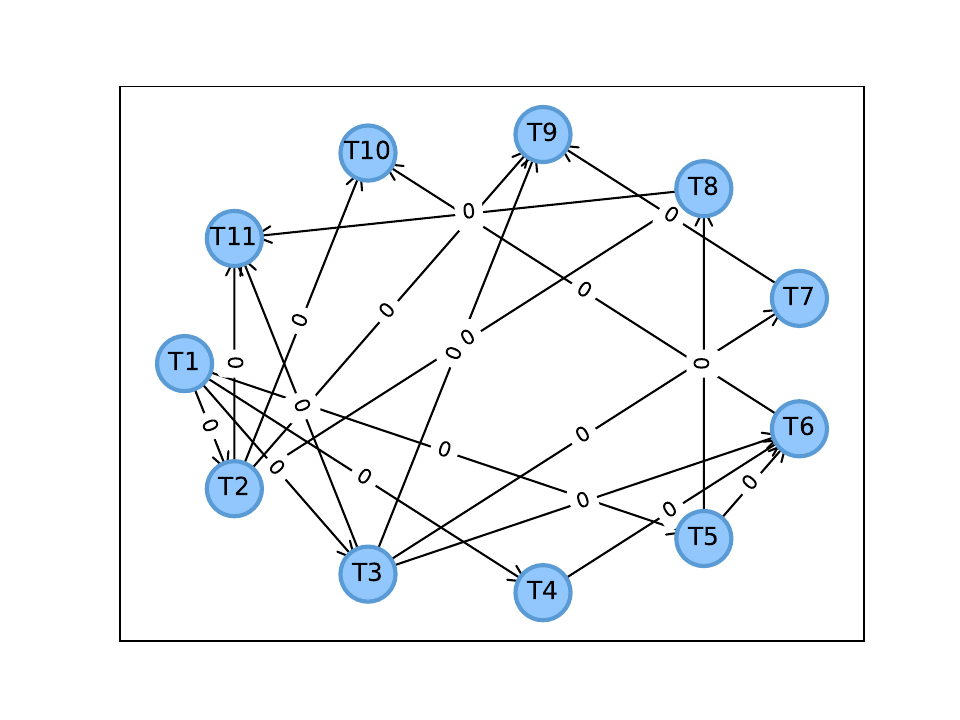}
        \caption{W5\_STGS1\_(3Nx11T): Workflow without DTT.}
        \label{fig:WithoutDataTransferTime}
    \end{subfigure}\hfill
    \begin{subfigure}[t]{0.32\textwidth}
        \centering        
        \includegraphics[width=\textwidth]{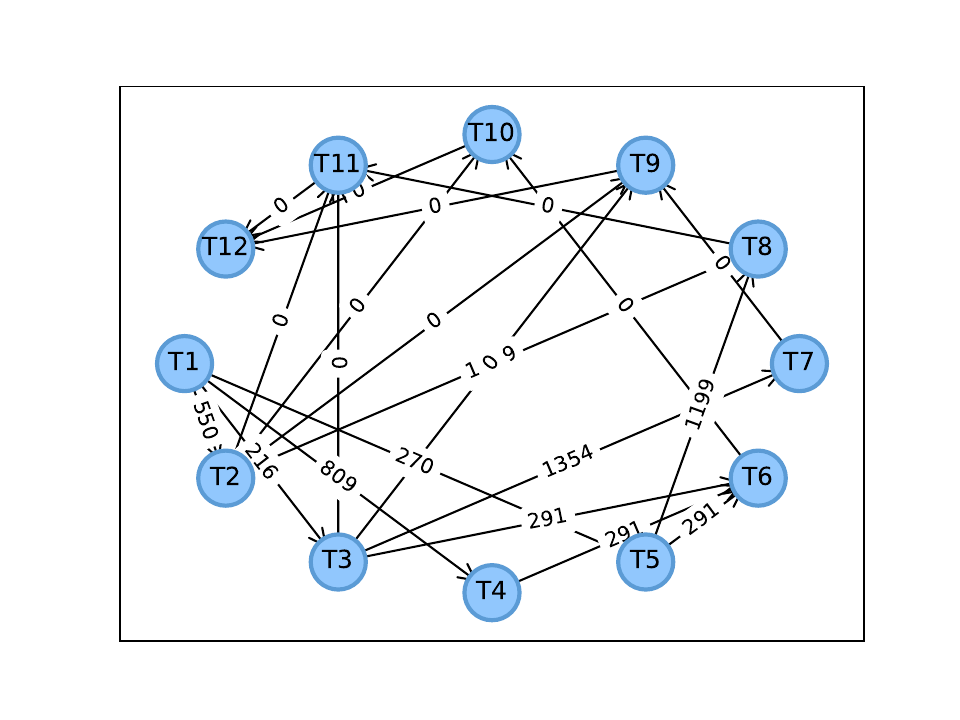}      
        \caption{W6\_STGS2\_(3Nx12T): Workflow with DTT.}
        \label{fig:WithDataTransferTime}
    \end{subfigure}\hfill
    \begin{subfigure}[t]{0.32\textwidth}
        \raggedleft
        \includegraphics[width=\textwidth]{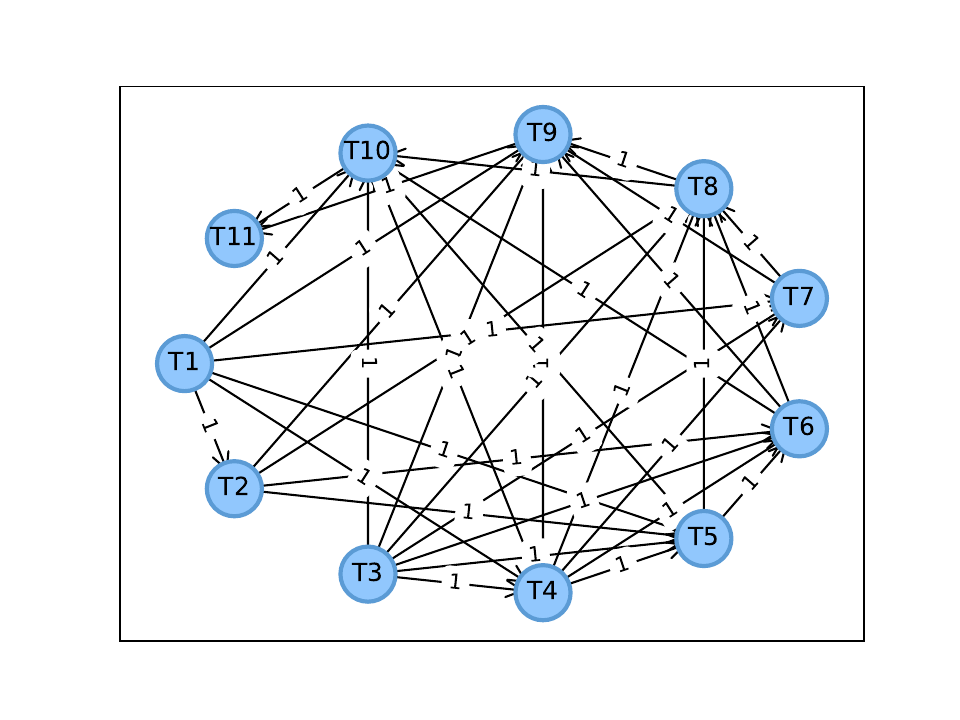}        
        \caption{W7\_STGS3\_(3Nx11T): Workflow with dense connections and default DTT.}
        \label{fig:WithProcessorCount}
    \end{subfigure}
    \caption{STGS Workflows.}
    \label{fig:STGSWorkflows}
    \vspace{-1em}
\end{figure*}

\begin{table}[ht]
    \caption{Workflow details for \textit{Test Case I} (\Cref{sssec:TestCaseQuality}).}
    \label{tab:TestCaseWorkflows}
    \centering
    \begin{tabular}{p{1em}|p{10em}|p{16em}}
        \hline
        \textbf{S.N.} & \textbf{Workflow Short Name} & \textbf{Full Form} \\
        \hline
        1. & W1\_Se\_(3Nx3T)   & MRI Serial Workflow \\
        2. & W2\_Pa\_(3Nx4T)   & MRI Parallel Workflow \\
        3. & W3\_Ra\_(3Nx5T)   & Random Workflow \\
        4. & W4\_Ra\_(3Nx10T)  & Random Workflow \\
        5. & W5\_STGS1\_(3Nx11T) & STGS Workflow Without Communication Cost \\
        6. & W6\_STGS2\_(3Nx12T) & STGS Workflow With Communication Cost \\
        7. & W7\_STGS3\_(3Nx11T) & STGS Workflow With Default Communication Cost \& Dense Connections \\
        \hline
    \end{tabular}
    \vspace{-1em}
\end{table}

\section{Experiments}

To evaluate the usability, we conduct a few more experiments, taking the realistic heterogeneous environments, varying workload intensities and node capabilities to evaluate performance and the scalability of the framework.
For this we assess it using real (from the Standard Task Graph Set \cite{tobita2002standard, kasaharaPracticalMultiprocessorScheduling1984a}) and synthetic workflows. 
We also consider workflows with varying speed and communication cost across different node types; such variations are critical in real-world settings as they can significantly affect overall execution times. That is why , the experiments are designed to assess the scheduler's adaptability to these variations. 

\subsubsection{Experiment Environment}
\label{sssec:TestSystem}
Experiment was conducted on an edge node (8 cores, 8 GB RAM) connected to the HPC-CC. A small edge node is taken to test the solver's capability under limited resources, we then use higher resources for scale tests. The solver is implemented using Python language.

\subsubsection{Test Case I - Performance Quality}
\label{sssec:TestCaseQuality}
\begin{figure}
    \centering
    \includegraphics[width=0.5\textwidth]{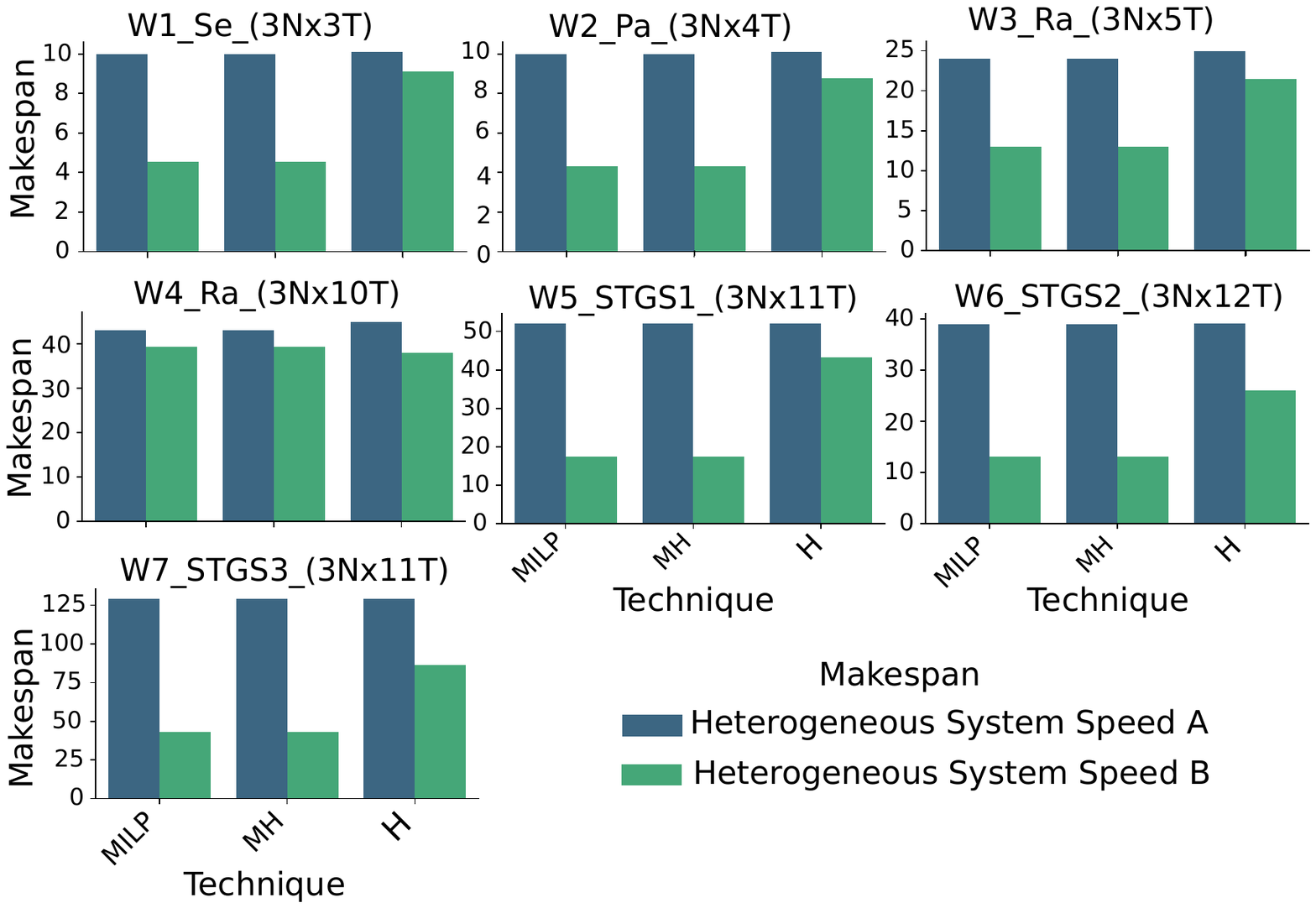}
    \caption{Faceted view of makespan variations for different techniques under varying processing speeds. (See Table~\ref{tab:TestCaseWorkflows} for abbreviations.)}
    \label{fig:QualityTest_FacetedViewGraph_Modified_V2}
    \vspace{-1em}
\end{figure}

Figure~\ref{fig:QualityTest_FacetedViewGraph_Modified_V2} shows the faceted view of makespan variations for different workflows (\Cref{fig:MRIWorkflow} \& \Cref{fig:STGSWorkflows}) under varying processing speeds ($A: 1\times$ and $B: 2\times$). Table~\ref{tab:TestCaseWorkflows} details the workflow abbreviations. The findings shows that the MILP consistently suggested the optimal makespan, whereas meta-heuristic (MH) and heuristic (H) techniques yield approximate (suboptimal) makespan but comparatively in shorter time.

\subsubsection{Test Case II - Scalability}
\label{sssec:TestCaseScalability}
Scalability tests were conducted using synthetic workflows with varying numbers of nodes and tasks characteristics. The findings in Table~\ref{tab:ScaleTest}, shos  that MILP does not scale well for large instances, while heuristic techniques are highly scalable then MILP and Meta-heuristic methods providing faster but sub-optimal solution.

\begin{table}[ht]
    \centering
    \caption{Scale Test}
    \begin{tabular}{cp{9em}r}
        \hline
        \textbf{Num Nodes x Tasks} & \textbf{Techniques} & \textbf{Time to Solution (sec)} \\       
        \hline
        5 x 5 & MILP & 0.02 \\ 
        5 x 5 & MH   & 0.03 \\ 
        5 x 5 & H    & 0.00 \\ 
        \hline
        50 x 50 & MILP & - \\ 
        50 x 50 & MH   & 77.78 \\ 
        50 x 50 & H    & 0.01 \\         
        \hline
        500 x 500 & MILP & - \\ 
        500 x 500 & MH   & 6513.33 \\ 
        500 x 500 & H    & 0.24 \\ 
        \hline
        5000 x 5000 & MILP & - \\ 
        5000 x 5000 & MH   & - \\ 
        5000 x 5000 & H    & 560.59 \\         
        \hline
    \end{tabular}
    \label{tab:ScaleTest}
    \vspace{-1em}
\end{table}

\subsection{Discussion}
\label{ssec:Discussion}
Our observations confirm that the proposed system and workload modeling effectively supports practical and dynamic resource allocation and scheduling in HPC-CC environments using effective tools and techniques. The inputs taken in JSON formats (Figures~\ref{fig:InputNodes} and \ref{fig:InputWorkflow}) are compatible with Snakemake’s configuration standards, enabling structured definition of system resources and workflows. Our extended scheduler with MILP, meta-heuristic, and heuristic techniques, could perform in dynamic and realistic setup providing the best optimal solution based on the incoming workloads.

Analytical and experimental validations further verifies the effectiveness of the solution framework. Furthermore, the quality (Figure~\ref{fig:QualityTest_FacetedViewGraph_Modified_V2}) and scalability experiments (Table~\ref{tab:ScaleTest}) verify the practical usages of the modeling framework across diverse systems and workloads in an online environment.

While our framework successfully integrates and enhance the existing tools and techniques (e.g., Snakemake, MILP solvers), but still there remain some limitations regarding dynamic resource constraints and the complexity in multiple constraints handling. Future work will focus on enhancing and improving the model adaptability and robustness for further complex use cases, thereby improving the scheduling efficiency and effectiveness in heterogeneous HPC-CC landscapes.

\section{Conclusion And Recommendations}
\label{sec:ConclusionAndRecommendations}

This study demonstrated detailed system and workload modeling, incorporating practical tools and techniques, respecting user-specified requirements, tools' parameters, and adaptive resource allocation and scheduling requirements in heterogeneous HPC compute continuum environments. Our findings highlight that a structured approach—integrating standardized JSON input formats, an extended Snakemake scheduler, applying an adaptive scheduling approach—significantly could automate and enhance the scheduling efficiency and resource utilization.

Through rigorous experimental evaluations, we observed that the MILP-based approach achieved 100\% optimality for small-scale workflows (5×5 tasks) in as little as 0.02 seconds, closely matching manual estimations (Table~\ref{tab:SolutionMIP}, Figure~\ref{fig:MIP_OptimumScheduleGanttChart}). However, as workflow complexity increased (e.g., 50×50 tasks and beyond), MILP struggled to scale efficiently. In contrast, meta-heuristic and heuristic methods reduced computation times by up to 99\% for large-scale cases (5000×5000 tasks), though with a 5–10\% deviation from optimal makespan (Figure~\ref{fig:QualityTest_FacetedViewGraph_Modified_V2}, Table~\ref{tab:ScaleTest}). These results confirm the trade-off between solution optimality and scalability, reinforcing the need for hybrid approaches.

Despite these promising outcomes, we identified limitations in handling dynamic resource constraints and the complexity of constraint management in heterogeneous systems. Future research should focus on integrating real-time monitoring data and adaptive scheduling mechanisms to improve responsiveness to system variability. Additionally, we recommend exploring hybrid scheduling strategies from emerging technologies like Artificial Intelligence and Quantum Constraints Programming to balance optimality and computational efficiency. These advancements will contribute to more resilient, scalable, and high-performance workload management solutions in the evolving HPC-CC landscape.

\section*{Acknowledgment}
\label{sec:Acknowledgment}
The DECICE Project from the University of Göttingen, primarily funded by the European Horizon Project (Grant No. 101092582), is committed to the development of a robust Digital Twin Platform as a Service (DT-PaaS).

We commend the significant influence exerted by the deeply dedicated contributors from the Gesellschaft für wissenschaftliche Datenverarbeitung mbH Göttingen (GWDG) in shaping this study.

Special thanks to the Federal Ministry of Education and Research and state governments (visible at www.nhr-verein.de/unsere-partner), whose unified backing and funding through the National High Performance Computing (NHR) have definitively propelled this project's trajectory towards success.


\begin{thebibliography}{10}

\bibitem{ApplyComputeProject}
{Gottingen} {University}: {Scientific} {Compute} {Cluster}.

\bibitem{GurobiOptimizer}
{Linear} {Programming} {Tool}: {Gurobi} {Optimizer}.

\bibitem{ORTools}
{Linear} {Programming} {Tool}: {OR-Tools} {Optimizer}.

\bibitem{OptimizationPuLPPuLP}
{Linear} {Programming} {Tool}: {PuLP} {Optimizer}.

\bibitem{openpbs_nodate}
{OpenPBS} - {An} {open} {source} {project} {on} {Portable} {Batch} {System}).

\bibitem{NeumannArchitecture2022}
{Von} {Neumann} {Architecture}.

\bibitem{WhatHighPerformanceComputing2022}
What {{Is High-Performance Computing}} ({{HPC}})? | {{IBM}}.

\bibitem{LinearProgrammingExtensions1998}
Linear {{Programming}} and {{Extensions}} | {{Princeton University Press}},
  Sun, 08/23/1998 - 12:00.

\bibitem{abouelyazidMachineLearningAlgorithms2021}
Mahmoud Abouelyazid.
\newblock Machine {{Learning Algorithms}} for {{Dynamic Resource Allocation}}
  in {{Cloud Computing}}: {{Optimization Techniques}} and {{Real-World
  Applications}}.
\newblock 1(2):1--58.

\bibitem{achterbergConstraintIntegerProgramming2008}
Tobias Achterberg, Timo Berthold, Thorsten Koch, and Kati Wolter.
\newblock Constraint {{Integer Programming}}: {{A New Approach}} to {{Integrate
  CP}} and {{MIP}}.
\newblock In Laurent Perron and Michael~A. Trick, editors, {\em Integration of
  {{AI}} and {{OR Techniques}} in {{Constraint Programming}} for
  {{Combinatorial Optimization Problems}}}, volume 5015, pages 6--20. Springer
  Berlin Heidelberg.

\bibitem{agarwalDataintensiveScienceTerapixel2011}
Deb Agarwal, You-Wei Cheah, Dan Fay, Jonathan Fay, Dean Guo, Tony Hey, Marty
  Humphrey, Keith Jackson, Jie Li, Christophe Poulain, Youngryel Ryu, and
  Catharine Van~Ingen.
\newblock Data-intensive science: {{The Terapixel}} and {{MODISAzure}}
  projects.
\newblock 25(3):304--316.

\bibitem{agarwalTopologyawareTaskMapping2006}
T.~Agarwal, A.~Sharma, A.~Laxmikant, and L.V. Kale.
\newblock Topology-aware task mapping for reducing communication contention on
  large parallel machines.
\newblock In {\em Proceedings 20th {{IEEE International Parallel}} \&
  {{Distributed Processing Symposium}}}, pages 10 pp.--.

\bibitem{alamResourceawareLoadBalancing2020}
Mahfooz Alam, Raza~Abbas Haidri, and Mohammad Shahid.
\newblock Resource-aware load balancing model for batch of tasks ({{BoT}}) with
  best fit migration policy on heterogeneous distributed computing systems.
\newblock 16(2):113--141.

\bibitem{alghamdiDynamicClusteringbasedTask2024a}
Mona Alghamdi, Atm Alam, Arumugam Nallanathan, and Asma Cherif.
\newblock Dynamic {{Clustering-based Task Orchestrator}} in {{Mobile Edge
  Computing}}.
\newblock In {\em 2024 {{International Wireless Communications}} and {{Mobile
  Computing}} ({{IWCMC}})}, pages 1613--1618.

\bibitem{arifApplicationAttunedMemoryManagement2024}
Moiz Arif, Avinash Maurya, M.~Mustafa Rafique, Dimitrios~S. Nikolopoulos, and
  Ali~R. Butt.
\newblock Application-{{Attuned Memory Management}} for {{Containerized HPC
  Workflows}}.
\newblock In {\em 2024 {{IEEE International Parallel}} and {{Distributed
  Processing Symposium}} ({{IPDPS}})}, pages 114--127.

\bibitem{bestuzhevaEnablingResearchSCIP2023}
Ksenia Bestuzheva, Mathieu Besançon, and et.~al. Chen.
\newblock Enabling {{Research}} through the {{SCIP Optimization Suite}} 8.0.
\newblock 49(2):1--21.

\bibitem{boothNUMAAwareVersionAdaptive2024}
Joshua~Dennis Booth and Phillip Lane.
\newblock A {{NUMA-Aware Version}} of an {{Adaptive Self-Scheduling Loop
  Scheduler}}.

\bibitem{burns2019kubernetes}
Brendan Burns, Brian Grant, David Oppenheimer, Eric Brewer, and John Wilkes.
\newblock {\em Kubernetes: Up and Running: Dive into the Future of
  Infrastructure}.
\newblock O'Reilly Media, 2019.

\bibitem{castro2019serverless}
P.~Castro, V.~Ishakian, V.~Muthusamy, and A.~Slominski.
\newblock The rise of serverless computing.
\newblock {\em Communications of the ACM}, 62(12):44--54, 2019.

\bibitem{delahayeSimulatedAnnealingBasics2019}
Daniel Delahaye, Supatcha Chaimatanan, and Marcel Mongeau.
\newblock Simulated {{Annealing}}: {{From Basics}} to {{Applications}}.
\newblock In Michel Gendreau and Jean-Yves Potvin, editors, {\em Handbook of
  {{Metaheuristics}}}, volume 272, pages 1--35. Springer International
  Publishing.

\bibitem{eadlineHighPerformanceComputing2009}
Douglas Eadline.
\newblock {\em High Performance Computing for {{Dummies}}: Learn to: Pick out
  Hardware and Software, Find the Best Vendor to Work with, Get Your People up
  to Speed on {{HPC}}}.
\newblock {Wiley}, {Hoboken, NJ}, 2009.

\bibitem{expositoPerformanceAnalysisHPC2013}
Roberto~R. Expósito, Guillermo~L. Taboada, Sabela Ramos, Juan Touriño, and
  Ramón Doallo.
\newblock Performance analysis of {{HPC}} applications in the cloud.
\newblock 29(1):218--229.

\bibitem{fangTaskSchedulingStrategy2022}
Juan Fang, Jiaxing Zhang, Shuaibing Lu, Hui Zhao, Di~Zhang, and Yuwen Cui.
\newblock Task {{Scheduling Strategy}} for {{Heterogeneous Multicore Systems}}.
\newblock 11(1):73--79.

\bibitem{fatehiEnergyAwareMulti2021}
Saeed Fatehi, Homayun Motameni, Behnam Barzegar, and Mehdi Golsorkhtabaramiri.
\newblock Energy {{Aware Multi Objective Algorithm}} for {{Task Scheduling}} on
  {{DVFS-Enabled Cloud Datacenters}} using {{Fuzzy NSGA-II}}.
\newblock 12(2):2303--2331.

\bibitem{frachtenbergJobSchedulingStrategies2007}
Eitan Frachtenberg and Uwe Schwiegelshohn, editors.
\newblock {\em Job {{Scheduling Strategies}} for {{Parallel Processing}}: 12th
  {{International Workshop}}, {{JSSPP}} 2006, {{Saint-Malo}}, {{France}},
  {{June}} 26, 2006, {{Revised Selected Papers}}}, volume 4376 of {\em Lecture
  {{Notes}} in {{Computer Science}}}.
\newblock {Springer}.

\bibitem{gadParticleSwarmOptimization2022}
Ahmed~G. Gad.
\newblock Particle {{Swarm Optimization Algorithm}} and {{Its Applications}}:
  {{A Systematic Review}}.
\newblock 29(5):2531--2561.

\bibitem{gadbanInvestigatingOverheadREST2020a}
Frank Gadban, Julian Kunkel, and Thomas Ludwig.
\newblock Investigating the {{Overhead}} of the {{REST Protocol When Using
  Cloud Services}} for {{HPC Storage}}.
\newblock In Heike Jagode, Hartwig Anzt, Guido Juckeland, and Hatem Ltaief,
  editors, {\em High {{Performance Computing}}}, volume 12321, pages 161--176.
  Springer International Publishing.

\bibitem{happeSelfadaptiveHeterogeneousMulticore2013}
Markus Happe, Enno Lübbers, and Marco Platzner.
\newblock A self-adaptive heterogeneous multi-core architecture for embedded
  real-time video object tracking.
\newblock 8(1):95--110.

\bibitem{jiangEnergyefficientSchedulingFlexible2022}
Xingyu Jiang, Zhiqiang Tian, and et.~al. Liu.
\newblock Energy-efficient scheduling of flexible job shops with complex
  processes: {{A}} case study for the aerospace industry complex components in
  {{China}}.
\newblock 27:100293.

\bibitem{kasaharaPracticalMultiprocessorScheduling1984a}
{Kasahara} and {Narita}.
\newblock Practical {{Multiprocessor Scheduling Algorithms}} for {{Efficient
  Parallel Processing}}.
\newblock C-33(11):1023--1029.

\bibitem{koster2012snakemake}
Johannes Koster and Sven Rahmann.
\newblock Snakemake--a scalable bioinformatics workflow engine.
\newblock {\em Bioinformatics}, 28(19):2520--2522, 2012.

\bibitem{Koster:2012}
Johannes K{\"o}ster and Sven Rahmann.
\newblock Snakemake--a scalable bioinformatics workflow engine.
\newblock {\em Bioinformatics}, 28(19):2520--2522, 2012.

\bibitem{kunkelSimulationParallelPrograms2013}
Julian~Martin Kunkel.
\newblock Simulation of {{Parallel Programs}} on {{Application}} and {{System
  Level}}.

\bibitem{KunzmanProgrammingHeterogeneousSystems2011}
David~M. Kunzman and Laxmikant~V. Kale.
\newblock Programming heterogeneous systems.
\newblock In {\em 2011 IEEE International Symposium on Parallel and Distributed
  Processing Workshops and Phd Forum}, pages 2061--2064, 2011.

\bibitem{liAdvancesTopologyAwareScheduling2017}
Kangkang Li, Maciej Malawski, Piotr Oleksy, and Jarek Nabrzyski.
\newblock Advances in {{Topology-Aware Scheduling}} in {{Multidimensional
  Torus-Based Systems}}.
\newblock In {\em New {{Frontiers}} in {{High Performance Computing}} and {{Big
  Data}}}, pages 93--118. IOS Press.

\bibitem{maOptimizedWorkflowScheduling2021}
Haoyang Ma and Juan Fang.
\newblock An optimized workflow scheduling algorithm on {{CPU-GPU}}
  heterogeneous systems.
\newblock 1994(1):012037.

\bibitem{mao2019learning}
H.~Mao, M.~Schwarzkopf, S.~Venkatakrishnan, Z.~Meng, and M.~Alizadeh.
\newblock Learning scheduling algorithms for data processing clusters.
\newblock {\em Proceedings of the ACM on Measurement and Analysis of Computing
  Systems}, 3(1):1--35, 2019.

\bibitem{mishraLoadBalancingCloud2020}
Sambit~Kumar Mishra, Bibhudatta Sahoo, and Priti~Paramita Parida.
\newblock Load balancing in cloud computing: {{A}} big picture.
\newblock 32(2):149--158.

\bibitem{nicohabermannDesktopTeraflopExploiting1993}
{Nico Habermann}.
\newblock From {{Desktop}} to {{Teraflop}}: {{Exploiting The U}}.{{S}}.
  {{Lead}} in {{High Performance Computing}}.

\bibitem{paganoMakingControlHigh2024}
Rosa Pagano, Sophie Cerf, Bogdan Robu, Quentin Guilloteau, Raphael Bleuse, and
  Eric Rutten.
\newblock Making {{Control}} in {{High Performance Computing}} for {{Overload
  Avoidance Adaptive}} in {{Time}} and {{Job Size}}.
\newblock In {\em 2024 {{IEEE Conference}} on {{Control Technology}} and
  {{Applications}} ({{CCTA}})}, pages 753--760.

\bibitem{pinchakPracticalHeterogeneousPlaceholder2002}
Christopher Pinchak, Paul Lu, and Mark Goldenberg.
\newblock Practical {{Heterogeneous Placeholder Scheduling}} in {{Overlay
  Metacomputers}}: {{Early Experiences}}.
\newblock In Dror~G. Feitelson, Larry Rudolph, and Uwe Schwiegelshohn, editors,
  {\em Job {{Scheduling Strategies}} for {{Parallel Processing}}}, Lecture
  {{Notes}} in {{Computer Science}}, pages 205--228. Springer.

\bibitem{singhAutonomicResourceManagement2021}
Bhupesh Singh, Mohammad Danish, Tanupriya Choudhury, and Durga Sharma.
\newblock Autonomic {{Resource Management}} in a {{Cloud-Based Infrastructure
  Environment}}.
\newblock pages 325--345.

\bibitem{tanashImprovingHPCSystem2019}
Mohammed Tanash, Brandon Dunn, Daniel Andresen, William Hsu, Huichen Yang, and
  Adedolapo Okanlawon.
\newblock Improving {{HPC System Performance}} by {{Predicting Job Resources}}
  via {{Supervised Machine Learning}}.
\newblock In {\em Practice and {{Experience}} in {{Advanced Research
  Computing}} 2019: {{Rise}} of the {{Machines}} (Learning)}, {{PEARC}} '19,
  pages 1--8. Association for Computing Machinery.

\bibitem{tobita2002standard}
Takao Tobita and Hironori Kasahara.
\newblock A standard task graph set for fair evaluation of multiprocessor
  scheduling algorithms.
\newblock {\em Journal of Scheduling}, 5(5):379--394, 2002.

\bibitem{vanderaalstWorkflowManagementModels2002}
Wil~M.P. Van Der~Aalst and Kees Van~Hee.
\newblock {\em Workflow {{Management}}: {{Models}}, {{Methods}}, and
  {{Systems}}}.
\newblock The MIT Press.

\bibitem{wangRESCAPEResourceEstimation2024}
Jinghao Wang, Guangzu Wang, Tianyu Wo, Xu~Wang, and Renyu Yang.
\newblock {{RESCAPE}}: {{A Resource Estimation System}} for {{Microservices}}
  with {{Graph Neural Network}} and {{Profile Engine}}.
\newblock In {\em 2024 {{IEEE International Conference}} on {{Joint Cloud
  Computing}} ({{JCC}})}, pages 37--44.

\bibitem{wangRLSchertHPCJob2021}
Qiqi Wang, Hongjie Zhang, Cheng Qu, Yu~Shen, Xiaohui Liu, and Jing Li.
\newblock {{RLSchert}}: {{An HPC Job Scheduler Using Deep Reinforcement
  Learning}} and {{Remaining Time Prediction}}.
\newblock 11(20):9448.

\bibitem{yoo2003slurm}
Andy~B. Yoo, Martine~A. Jette, and Mark Grondona.
\newblock {SLURM}: Simple linux utility for resource management.
\newblock In {\em Proceedings of the 9th Workshop on Job Scheduling Strategies
  for Parallel Processing (JSSPP)}, pages 44--60. Springer, 2003.

\end{thebibliography}

\appendix

\end{document}